\documentclass[12pt]{iopart}

\usepackage{graphicx}
\usepackage{dcolumn}
\usepackage{bm}

\begin{document}

\title[Beltrami structures in disk-jet system]{Beltrami structures in disk-jet system: alignment of flow and generalized vorticity}

\author{Z. Yoshida$^{1}$ and N. L. Shatashvili$^{2,3}$}

\address{
$^{1}$Graduate School of Frontier Sciences, The University of Tokyo, Chiba 277-8561, Japan\\
$^{2}$Faculty of Exact and Natural Sciences, Javakhishvili Tbilisi State University, Tbilisi 0128, Georgia\\
$^{3}$Andronikashvili Institute of Physics, Javakhishvili Tbilisi
State University, Tbilisi 0177, Georgia}

\ead{yoshida@k.u-tokyo.ac.jp, shatash@ictp.it}

\begin{abstract}
The combination of a thin disk and a narrowly-collimated jet is a
typical structure that is observed in
the vicinity of a massive object such as AGN, black hole or YSO.
Despite a large variety of their scales and possible diversity of
involved processes, a simple and universal principle dictates the
geometric similarity of the structure; we show that the
singularity at the origin ($r=0$) of the Keplerian rotation
($V_\theta \propto r^{-1/2}$) is the determinant of the structure.
The collimation of jet is the consequence of the alignment
---so-called Beltrami condition--- of the flow velocity and the
``generalized vorticity'' that appears as an axle penetrating the
disk (the vorticity is generalized to combine with magnetic field
as well as to subtract the friction force causing the accretion).
Typical distributions of the density and flow velocity are
delineated by a similarity solution of the simplified version of
the model.
\end{abstract}

\maketitle


\section{Introduction}

An accretion disk often combines with spindle-like jet of ejecting
gas, and constitutes a typical structure
that accompanies a massive object of various scales, ranging from
young stars \cite{bib:Hartigan} to galactic nuclei
\cite{bib:Jones}. The mechanism that rules each part of different
systems might be not universal. Since early 70s after the
discovery of radio galaxies and quasars (see e.g.
\cite{bib:begelman3} and references therein) the main evidence from
detecting jets in different classes of astrophysical systems which
are observed to produce collimated jets near the massive central
object prove the direct association with an accretion disk,
although possibly reflecting different accretion regimes; the
opposite is not true in some objects for which the accretion disks
do not require collimated jets (viscous transport/disk winds may
play the similar role in the energy balance)
\cite{bib:bland1,bib:livio,bib:ferrari}. The macroscopic disk-jet
geometry, however, bears a marked similarity despite the huge
variety of the scaling parameters such as Lorentz factor, Reynolds
number, Lundquist number, ionization fractions, etc.
\cite{bib:bland1,bib:bland1-2,bib:livio,bib:livio2,bib:ferrari,bib:zanni}.
The search for a general principle that dictates such similarity
provided the stimulus for this paper.

Before formulating a model of the global structure, we start by
reviewing the basic properties of disk-jet systems. In the disk
region, the transport processes of mass, momentum and energy
depend strongly on the scaling parameters; many different
mechanisms have been proposed and examined carefully. The
classical (collisional) processes are evidently insufficient to
account for the accretion rate, thus turbulent transports
(involving magnetic perturbations) must be
invoked\,\cite{bib:BalbusHawley}. Winds may also remove the
angular momentum from the
disk\,\cite{bib:magn,bib:pellet,bib:bogov}. The connection of the
disk and the jet is more complicated. While it is evident that the
mass and energy of the jet are fed by the accreting flow, the
mechanism and process of mass/energy transfer are still not clear.
It is believed that the major constituent of jets is the
material of an accretion disk surrounding the central object.
Although for the fastest outflows the contributions to the total
mass flux may come from outer regions as well \cite{bib:zanni}. In
AGN one may think of taking some energy from the central black
hole.  Livio\,\cite{bib:livio} summarizes the conjectures as
follows: (i) powerful jets are produced by systems in which on top
of an accretion disk threaded by a vertical field, there exists an
additional source of energy/wind, possibly associated with the
central object (for example, stellar wind from porotostar may
accelerate YSO jets, as estimated by
Refs\,\cite{bib:ferreira,bib:ferreira2,bib:ferreira3}); (ii)
launching of an outflow from an accretion disk requires a hot
corona or a supply from some additional source of energy (see also
\cite{bib:matsumoto,bib:matsumoto2,bib:matsumoto3}); (iii)
extensive hot atmosphere around the compact object can provide
additional acceleration. Magnetic fields are considered to play an
important role in defining the local
accretion\,\cite{bib:bland1,bib:bland1-2,bib:bland2,bib:bland3,bib:bland5,bib:zanni}.
When magnetic field is advected inwards by accreting material
or/and generated locally by some mechanism, i.e. if there is a
bundle of magnetic field lines which is narrowest near the origin
(central object) and broader afar off, the centrifugal force due
to rotation may boost the jet along the magnetic field lines up to
a super-Alfv\'enic
speed\,\cite{bib:begelman3,bib:begelman4,bib:begelman5,bib:bland4,bib:shibata,bib:shibata2,bib:anderson2}.
In the case of AGN, there is an alternative idea suggested by
Blandford \& Znajek\,\cite{bib:bland1} based on electro-dynamical
processes extracting energy from a rotating black hole.
Extra-galactic radio jets might be accelerated by highly
disorganized magnetic fields that are strong enough to dominate
the dynamics until the terminal Lorentz factor is
reached\,\cite{bib:begelman-pressure}. Following the twin-exhaust
model by Blandford \& Rees\,\cite{bib:bland}, the collimation
under this scenario is provided by the stratified thermal pressure
from an external medium. The acceleration efficiency then depends
on the pressure gradient of the medium. In addition to the
energetics to account for the acceleration of ejecting flow, we
have to explain how the streamlines/magnetic-field lines change
the topology through the disk-jet connection\,\cite{bib:Shiraishi}
(see also \cite{bib:shakura,bib:bland1-2}).

Despite the diversity and complexity of holistic processes, there
must yet be a simple and universal principle that determines the
geometric similarity of disk-jet compositions. In the present
paper we will show that the collimated structure of jet is a
natural consequence of the \emph{alignment} of the velocity and
the ``generalized vorticity'' ---here the notion of \emph{vorticity}
will be generalized to combine with magnetic field (or, electromagnetic
vorticity) as well as to include the effective friction causing
the accretion. In a Keplerian thin disk, the vorticity becomes
a vertical vector with the norm $\propto r^{-3/2}$ ($r$ is the
radius from the center of the disk), which appears as a spindle of
a thin disk. Then, \emph{alignment} is the only recourse in
avoiding singularity of force near the axis of Keplerian rotation.
We are not going to discuss the ``process'' that produces the
global structure; we will, however, elucidate a ``necessary
condition'' imposed on the structure that can live long.

To demonstrate how the alignment condition arises and how it
determines the \emph{singular} structure of a \emph{thin} disk and
\emph{narrowly-collimated} jet, we invoke the simplest (minimum)
model of magnetohydrodynamics (Sec.\,\ref{sec:formulation}). The
alignment condition (so-called \emph{Beltrami} relation) will be
derived in Sec.\,\ref{Sec:Disk-Jet}. In Sec.\,\ref{sec:Beltrami},
we will formulate a system of equations which gives general
\emph{Beltrami structures} satisfying the alignment condition. In
Sec.\,\ref{sec:similarity-solution}, we will give an analytic
solution to a simplified version of the equations by a
similarity-solution method.

\section{Momentum Equation}
\label{sec:formulation}

We start by formulating a magnetohydrodynamic (MHD)
model. Let $ \bm{P} = \rho \bm{V}$ denote the momentum density,
where $\rho$ is the mass density and $\bm{V}$ is the (ion) flow
velocity. The momentum balance equations for the ion and electron
fluids are
\begin{eqnarray}
& &
\partial_t \bm{P} + \nabla\cdot(\rho\bm{V}\otimes\bm{V})
\nonumber \\
& &~~~~= \epsilon^{-1}\rho\left(\bm{V}\times\bm{B}-\partial_t\bm{A}\right)
\nonumber \\
& &~~~~~~~~~-\rho\nabla(\phi+\varphi) - \nabla p_i - \nu \bm{P},~~~~
\label{momentum-i-1}
\\
& &~~~0= \epsilon^{-1}\rho\left(\bm{V}_e \times\bm{B} -\partial_t\bm{A}\right)
-\rho\nabla\varphi +\nabla p_e,~~~~
\label{momentum-e-1}
\end{eqnarray}
where $\bm{B}$ is the magnetic field, $\varphi$ and $\bm{A}$ are
the the electromagnetic potentials, $\phi$ is the gravity
potential, $p_i$ and $p_e$ are the ion and electron pressures, and
$\nu$ is the (effective) friction coefficient.
We are assuming singly charged ions, for
simplicity. The variables are normalized as follows: We choose a
representative flow velocity $V_0$ and a mass density $\rho_0$ in
the disk, and normalize $\bm{V}$ and $\rho$ by these units. The
energy densities \ $|\bm{B}|^2/8\pi$ , \ $e\varphi$, \ $\rho\phi$,
\ $p_i$ and $p_e$ are normalized by the unit kinetic energy
density
\begin{equation}
{\cal{E}}_0=\frac{\rho_0 V_0^2}{2}. \label{Unit_energy}
\end{equation}
The independent variables (coordinate $\bm{x}$ and time $t$) are
normalized by the system size $L_0$ and the corresponding transit
time $T_0=L_0/V_0$. The scale parameter $\epsilon$ is defined by
\[
\epsilon = \frac{\delta_i}{L_0},
\]
where $\delta_i = mc/\sqrt{4\pi e^2\rho_0}$ is the ion inertia length
($m$: ion mass, $e$: elementary charge).

In equation\,(\ref{momentum-e-1}), we are neglecting the inertia and
the gravitation of electrons, as well as the electric resistivity.
The electron velocity $\bm{V}_e$ is given by
\[
\bm{V}_e = \bm{V}- \epsilon \rho^{-1}\nabla\times\bm{B} .
\]
The system (\ref{momentum-i-1})-(\ref{momentum-e-1}) includes a
finite dissipation by the effective friction force on the (ion)
momentum, not by conventional viscosities or resistivity. This is
for the simplicity of analysis. We do need a finite dissipation to
allow accretion, but the detail mechanism of dissipation is not
essential in the scope of our arguments.

In this paper, we consider stationary solutions, so we set
$\partial_t=0$. To study the macroscopic structure of an
astronomical system, we may assume $\epsilon \ll 1$.  Let us
expand $\bm{B}$ as
\[
\bm{B} = \bm{B}^{(0)} + \epsilon \bm{B}^{(1)} + \cdots,
\]
where the magnitude of each $\bm{B}^{(n)}$ is of order unity.
All other independent and dependent variables are assumed to be at
most of order unity. Then,
equation\,(\ref{momentum-e-1}) reads as
\begin{eqnarray*}
0&=&\epsilon^{-1}\rho\bm{V} \times\bm{B}^{(0)}
\\ & &
+\rho\left[
\bm{V}\times\bm{B}^{(1)}-\rho^{-1}(\nabla\times\bm{B}^{(0)})\times\bm{B}^{(0)}\right]
\\ & &
-\rho\nabla\varphi +\nabla p_e 
+ O(\epsilon).
\end{eqnarray*}
The term of order $\epsilon^{-1}$ demands
\begin{equation}
\bm{B}^{(0)} = \mu \bm{P}, \label{Beltrami-0}
\end{equation}
where $\mu$ is a certain scaler function,
which is the reciprocal Alfv\'en Much number.
Operating divergence on
both sides of equation (\ref{Beltrami-0}), we find $\nabla\cdot(\mu \bm{P}) =
\bm{P}\cdot\nabla\mu=0$. From order $\epsilon^{0}$ terms, we
obtain
\[
\bm{V}\times\bm{B}^{(1)} = \rho^{-1}(\nabla\times\bm{B}^{(0)})\times\bm{B}^{(0)}
+\nabla\varphi - \rho^{-1}\nabla p_e. 
\]
Using these relations in equation\,(\ref{momentum-i-1}),
we obtain
\begin{equation}
\nabla\cdot(\rho\bm{V}\otimes\bm{V}) = [\nabla\times(\mu\bm{P})]\times(\mu\bm{P})
-\rho\nabla\phi - \nabla p - \nu \bm{P}, \label{momentum-i-1-scaled}
\end{equation}
where $p=p_i+p_e$.

In order to derive a term that balances with the friction term, we
decompose the ``inertia term'' [the left-hand side of
equation\,(\ref{momentum-i-1-scaled})] as follows:  we first write
\begin{equation}
\rho = \rho_1 \rho_2  \label{decomposition1}
\end{equation}
($\rho_2$ will be determined as a function of $\nu$ in equation\,(\ref{friction-balance})),
and denote
\begin{equation}
\bm{{P}_1} = \rho_1\bm{V}, \quad \bm{P}_2 = \rho_2 \bm{V}. \label{decomposition2}
\end{equation}
Using these variables, we may write
\begin{eqnarray}
\nabla\cdot(\rho\bm{V}\otimes\bm{V})&=&\nabla\cdot(\bm{P}_1\otimes\bm{P}_2)
\nonumber \\
&=&(\nabla\cdot\bm{P}_1)\bm{P}_2 + (\bm{P}_1\cdot\nabla)\bm{P}_2 .~~
 \label{decomposition3}
\end{eqnarray}
In the conventional formulation of fluid mechanics,
we choose $\rho_2=1$ and $\rho_1=\rho$.  Then,
$\nabla\cdot(\rho\bm{V}\otimes\bm{V}) = (\nabla\cdot\bm{P})\bm{V}
+ \rho(\bm{V}\cdot\nabla)\bm{V}$. Combining with $\partial_t
\bm{P}=\rho\,\partial_t\bm{V} + \bm{V}\,\partial_t \rho$, and
using the mass conservation law $\partial_t \rho + \nabla\cdot\bm{P}=0$,
we observe that the left-hand side of
equation\,(\ref{momentum-i-1-scaled}) reduces into the standard inertia
term $\rho(\bm{V}\cdot\nabla)\bm{V}$ (here we are considering
the steady state with $\partial_t\bm{V} =0$). In the preset
analysis, however, we choose a different separation of the inertia
term to match the term $(\nabla\cdot\bm{P}_1)\bm{P}_2$ with the
friction term $-\nu\bm{P}$; this strategy will be explained in
the next section.

Multiplying $(\rho_2/\rho_1)$ on both sides of
equation\,(\ref{momentum-i-1-scaled}), we obtain
\begin{eqnarray}
\bm{P}_2\times\bm{\Omega} &=& \frac{1}{2}\nabla P_2^2 + \rho_2^2 \nabla\,(\phi + h)
\nonumber \\
& & +\rho_2\nu\,\bm{P}_2 + \frac{\rho_2}{\rho_1}\,(\nabla\cdot\bm{P}_1)\bm{P}_2 ,
\label{momentum-2}
\end{eqnarray}
where
\begin{eqnarray}
\bm{\Omega} &=& \nabla\times\bm{P}_2 -
\mu\rho_2\nabla\times(\mu\,\bm{P})
\label{generalized-vorticity}
\\
&=& \nabla\times\bm{P}_2
- \mu\rho_2\nabla\times\bm{B}^{(0)}
\nonumber
\end{eqnarray}
is a \emph{generalized vorticity},
and $h$ is the enthalpy density ($\nabla h = \rho^{-1}\nabla p$).

\section{Disk-Jet Structure}
\label{Sec:Disk-Jet}

Here we assume toroidal symmetry ($\partial_\theta =0$ in the
$r$-$\theta$-$z$ coordinates). We consider a massive central object
(a singularity at the origin), which yields $\phi = -MG/r$ (we may
neglect the mass in the disk in evaluating $\phi$). Then, the flow
velocity is $\bm{V} \approx V_\theta \bm{e}_\theta$ with the
Keplerian velocity $V_\theta \propto r^{-1/2}$ in the disk region.
The vorticity is $\nabla\times\bm{V} = \Omega_z \bm{e}_z$ with
$\Omega_z \propto r^{-3/2}$.  The momentum is strongly localized
in the thin disk, and the vorticity diverges near the axis (in this macroscopic view).
This singular configuration allows only a special geometric structure
to emerge. By equation\,(\ref{momentum-2}), following conclusions are
readily deducible:

(i) In the disk, a radial flow (which is much smaller than $V_\theta$)
is caused by the friction. The friction force $\rho_2\nu
\,\bm{P}_2$ is primarily in the azimuthal (toroidal) direction,
which may be balanced with the term
$(\rho_2/\rho_1)\,(\nabla\cdot\bm{P}_1)\bm{P}_2$ that has been
extracted from the inertia term
$\nabla\cdot(\rho\bm{V}\otimes\bm{V})$; see
equation\,(\ref{decomposition3}).
Using the steady-state mass conservation law
$\nabla\cdot\bm{P}=0$, we observe
\begin{eqnarray*}
\rho_1^{-1} \nabla\cdot\bm{P}_1 &=& \rho_1^{-1} \nabla\cdot(\bm{P}/\rho_2)
\\
&=&  \rho_2 \bm{V}\cdot\nabla\rho_2^{-1} = - \bm{V}\cdot\nabla\log\rho_2 .
\end{eqnarray*}
Hence, the  balance of the friction and the partial inertia term demands
\begin{equation}
\bm{V}\cdot\nabla\log\rho_2 = \nu, \label{friction-balance}
\end{equation}
which determines the parameter $\rho_2$. The remaining part
$\rho_1$ of the density is, then, determined by the mass conservation
law: From
\begin{eqnarray*}
\nabla\cdot\bm{P}&=&\nabla\cdot(\rho\bm{V})
\\
&=& \rho_2\bm{V}\cdot\nabla\rho_1
+ \rho_1\bm{V}\cdot\nabla\rho_2 +
\rho_1\rho_2\nabla\cdot\bm{V}
\\
&=& 0
\end{eqnarray*}
and equation\,(\ref{friction-balance}), we obtain a relation
\begin{equation}
\bm{V}\cdot\nabla\log\rho_1 = -\nabla\cdot\bm{V}-\nu.
\label{friction-balance'}
\end{equation}

(ii)
After balancing the third and fourth terms in
equation\,(\ref{momentum-2}), the remaining terms must not have an
azimuthal (toroidal) component.  In fact, the right-hand side
(gradients) have only poloidal ($r$-$z$ plane) components, and
hence, the left-hand side cannot have a toroidal component.

(iii)
In the vicinity of the axis threading the central object, the flow
$\bm{V}(=\bm{P}_2/\rho_2$) must \emph{align} to the \emph{generalized vorticity}
$\bm{\Omega}$ (that is dominated by $\nabla\times\bm{V} \propto
r^{-3/2}\bm{e}_z$), producing the collimated jet structure, to
minimize the Coriolis force $\bm{V}\times\bm{\Omega}$:
Otherwise, the remaining potential forces (gradients of potentials
which have only poloidal components) cannot balance with the
rotational (toroidal) Coriolis force. The alignment condition,
which we call \emph{Beltrami condition}, reads as
\begin{equation}
\bm{\Omega} = \lambda \bm{P}, \label{Beltrami1}
\end{equation}
where $\lambda$ is a certain scalar function. The pure
electromagnetic Beltrami condition is the well-known
\emph{force-free} condition $\nabla\times\bm{B} =
\lambda\bm{B}$\,\cite{bib:chandrasekhar,bib:low,bib:yoshida-giga}.
Equation (\ref{Beltrami1}) demands the alignment of the
generalized vorticity and momentum\,\cite{bib:beltrami}.

(iv)
When the Beltrami condition eliminates the left-hand side of
equation\,(\ref{momentum-2}), the remaining potential forces must
balance and achieve the \emph{Bernoulli
condition}\,\cite{bib:beltrami} which reads as
\begin{eqnarray}
\frac{1}{2\rho_2^2}\nabla P_2^2 + \nabla( \phi + h)
&=& \nabla \left( \frac{1}{2}V^2 + \phi + h \right) + V^2 \nabla \log\rho_2
\nonumber \\
&=& 0.
\label{Bernoulli-1}
\end{eqnarray}


The system of determining equations is summarized as follows: By
equation\,(\ref{friction-balance}), we determine the ``artificial
ingredient'' $\rho_2$
for a given $\nu$. This equation involves $\bm{V}=\bm{P}/\rho$
which is governed by the Beltrami equation (\ref{Beltrami1}).
After determining $\bm{V}$ and $\rho_2$, we can solve the
Bernoulli equation (\ref{Bernoulli-1}) to determine the enthalpy
$h$ (the gravitational potential is approximated by $\phi=-MG/r$).

\section{Beltrami Vortex Structure}
\label{sec:Beltrami}

\subsection{General two-dimensional structure}
\label{subsec:general2D}

In this section, we rewrite the Beltrami-Bernoulli equations
(\ref{friction-balance})-(\ref{Bernoulli-1}) in a more
manageable form by invoking the Clebsch parameterization\,\cite{bib:yoshida2009}.
In an axisymmetric geometry, the divergence-free vector $\bm{P}$
may be parameterized as
\begin{equation}
\bm{P} = \nabla\psi\times\nabla\theta + I \nabla \theta,
\label{Grad-form-1}
\end{equation}
where $I = \rho r V_\theta$.  Both $\psi$ and $I$ do not depend on
$\theta$. Since $\bm{P}\cdot\nabla\psi=0$, the level sets
(contours) of $\psi$ are the streamlines of $\bm{P}$ (or those of
$\bm{V}=\bm{P}/\rho$).

Using the expression (\ref{Grad-form-1}) in
equation\,(\ref{friction-balance}) yields
\begin{eqnarray}
\nu = \frac{1}{\rho} \bm{P}\cdot\nabla \log \rho_2 &=&
\frac{1}{\rho}\nabla\log\rho_2\times\nabla\psi\cdot\nabla\theta
\nonumber
\\
&\equiv& \frac{1}{r\rho}\{\log\rho_2,\psi\},
\label{friction-balance-2}
\end{eqnarray}
where $\{a,b\} \equiv (\partial_r b) (\partial_z a) - (\partial_r
a)(\partial_z b)$. For a given set of $\bm{P}$, $\rho$ and $\nu$,
we can solve equation\,(\ref{friction-balance-2}) to determine $\rho_2$,
as well as $\rho_1=\rho/\rho_2$ which is consistent to
equation\,(\ref{friction-balance'}).

Let us rewrite the momentum equation (\ref{momentum-2}) using
the Clebsch parameterization (\ref{Grad-form-1}). We observe
\begin{eqnarray}
\bm{\Omega} &=& \nabla\times\left(\rho_1^{-1}\bm{P}\right) -
\mu\rho_2\nabla\times\left(\mu\bm{P}\right)
\nonumber
\\
&=& -\left[ \left(\rho_1^{-1}-\mu^2\rho_2 \right){\cal L}\psi
+ \nabla\psi\cdot\left(\nabla\rho_1^{-1} -\mu\rho_2\nabla\mu \right) \right] \nabla\theta
\nonumber \\
& &
+ \left[ \nabla \left(\rho_1^{-1}I \right) - \mu\rho_2
\nabla \left( \mu I \right) \right] \times \nabla \theta ,~
\label{Grad-form-2}
\end{eqnarray}
where
\[
{\cal L}\psi \equiv r \partial_r(r^{-1}\partial_r \psi) + \partial_z^2 \psi.
\]
As mentioned above, $\bm{P}_2\times\bm{\Omega}$ may not have a
toroidal component, i.e.,
\[
\left[ \nabla \left(\rho_1^{-1}I \right) - \mu\rho_2
\nabla \left( \mu I \right) \right] \times \nabla\psi =0,
\]
which is equivalent to the existence of a scalar function $\lambda$ such that
\begin{eqnarray}
\lambda \nabla\psi&=&
\nabla \left(\rho_1^{-1}I \right) - \mu\rho_2
\nabla \left( \mu I \right)
\nonumber \\
&=&
(\rho_1^{-1} - \mu^2\rho_2)\nabla I
+ I \nabla \rho_1^{-1}
- \mu\rho_2 I \nabla\mu.
\label{Beltrami-I}
\end{eqnarray}
The poloidal component of the momentum equation (\ref{momentum-2}) reads
\begin{eqnarray}
& & \left[ \left(\rho_1^{-1}-\mu^2\rho_2 \right)({\cal L}\psi)
\right.
\nonumber \\
& & ~~\left. + \nabla\psi\cdot\left(\nabla\rho_1^{-1} -\mu\rho_2\nabla\mu \right) \right]\nabla\psi
+ I \nabla I
\nonumber
\\
& & ~~~~
= - \frac{r^2}{2}\nabla \left[r^{-2}\rho_1^{-2}(|\nabla\psi|^2+ I^2) \right]
\nonumber
\\
& & ~~~~~~~~-r^2 \rho_2^2 \nabla(\phi + h) .
\label{Grad-form-3}
\end{eqnarray}

We have to determine a self-consistent set of functions $\psi$,
$I$, $h$, $\rho_1$, $\rho_2$, $\lambda$ and $\mu$ (the friction
coefficient $\nu$ and the gravitational potential $\phi=-MG/r$ are
given functions). Our system of equations are
(\ref{friction-balance-2}), (\ref{Beltrami-I}) and
(\ref{Grad-form-3}), which are simultaneous nonlinear partial
differential equations including hyperbolic parts. Since
$\bm{P}\cdot\nabla\mu = 0$ [see equation\,(\ref{Beltrami-0})], we may
assume that $\mu = \mu(\psi)$, and give this function as a
``Cauchy data'' (an arbitrary function of $\psi$ that is constant
along the characteristics). Another function (for example
$\rho_1$) may be arbitrarily prescribed to leave five unknown
functions to be determined by five equations of the system.
However, the hyperbolic parts must be integrated by some Cauchy
data, not by boundary values. We note that the factor
$(\rho_1^{-1}-\mu^2\rho_2)$ multiplying the elliptic operator
${\cal L}$ in equation\,(\ref{Grad-form-3}) may cause the \emph{Alfv\'en
singularity}.

To proceed with analytic calculations,
we will consider simplified systems in which we can
integrate the hyperbolic part of the equations easily.

\subsection{Unmagnetized Beltrami flow}

Here, we consider unmagnetized (or super-Alfv\'enic;
$\mu\approx0$) Beltrami-Bernoulli flows.
With $\mu=0$, equation\,(\ref{Beltrami-I}) reduces to
\begin{equation}
\lambda \nabla\psi =
\nabla \left(\rho_1^{-1}I \right) ,
\label{Beltrami-I-unmag}
\end{equation}
which implies $\rho_1^{-1}I = I_2(\psi)$ (Cauchy data) and
$\lambda = \lambda(\psi) = I_2'(\psi) $ (we denote
$f'(\psi)=df(\psi)/d\psi$). We notice that
equation\,(\ref{Beltrami-I-unmag}) is just the toroidal (azimuthal)
component of the Beltrami condition (\ref{Beltrami1}). Applying
the Beltrami condition also to the poloidal component of the
momentum equation, we can simplify equation\,(\ref{Grad-form-3}); with
$\mu=0$ and $I = \rho_1 I_2(\psi)$, we obtain
\begin{equation}
{\cal L} \psi
- \nabla\psi\cdot\nabla\log\rho_1
= - \rho_1^2 I_2'(\psi) I_2(\psi) .
\label{Beltrami3'}
\end{equation}
The Beltrami condition has decoupled the gradient forces [the
right-hand side of equation\,(\ref{Grad-form-3})] from the momentum
equation, which must balance separately: This is the Bernoulli
condition (\ref{Bernoulli-1}) which now reads as
\begin{eqnarray}
\nabla h &=&
-\nabla \left[ \frac{1}{2r^2\rho^2}\left(|\psi|^2+ I^2 \right) + \phi \right]
\nonumber \\
& & -  \frac{1}{r^2\rho^2}\left(|\psi|^2+ I^2 \right) \nabla \log\rho_2.
\label{Bernoulli-2}
\end{eqnarray}

Now the determining equations are much simplified --- for a given
distribution $\rho_1$ and Cauchy data $I_2(\psi)$, 
we may solve equation\,(\ref{Beltrami3'}) to determine $\psi$. Then, for
a given $\nu$, equation\,(\ref{friction-balance-2}) yields $\rho_2$.
Finally, the enthalpy $h$ is determined by equation\,(\ref{Bernoulli-2}).

\subsection{A simple magnetized Beltrami flow}
If we assume
\begin{equation}
\rho_1 = \rho_1(\psi),
\label{Beltrami-rho1}
\end{equation}
(i.e., $\bm{V}\cdot\nabla\rho_1 = 0$ and thus $\nabla\cdot
\bm{P}_2=0$), we may rather easily include a magnetic field into
the solution ($\mu=\mu(\psi)\neq 0$). By equation\,(\ref{friction-balance'}),
this assumption implies that the
compressibility is only by the frictional deceleration (then,
$0=\nabla\cdot\bm{P} = \nabla\cdot(\rho_1 \bm{P}_2) =
\nabla\cdot\bm{P}_2$).

Under the assumption (\ref{Beltrami-rho1}), the second and third
terms of equation\,(\ref{Beltrami-I}) parallel $\nabla\psi$. Therefore,
$\nabla I$ must align to $\nabla\psi$, implying $I = I(\psi)$. In
this case, $\lambda$ is given by
\begin{equation}
\lambda = (\rho_1^{-1} - \mu^2\rho_2)I' - (\rho_1^{-2}\rho_1' + \rho_2 \mu\mu')I .
\label{Beltrami-I-2}
\end{equation}
On the right-hand side of equation\,(\ref{Beltrami-I-2}), only $\rho_2$
is not a function of $\psi$.
The Beltrami condition demands vanishing of the left-hand side of
equation\,(\ref{Grad-form-3}), which, with a Cauchy data $I(\psi)$ and
the $\lambda$ of equation\,(\ref{Beltrami-I-2}), reads as
\begin{equation}
(\rho_1^{-1}-\mu^2\rho_2) {\cal L} \psi
- (\rho_1^{-2}\rho_1' + \rho_2 \mu\mu') |\nabla\psi|^2
= - \lambda I(\psi) .
\label{Beltrami3}
\end{equation}
This equation (governing $\psi$) is coupled with
equation\,(\ref{friction-balance-2}) through $\rho_2$. The Bernoulli
condition (\ref{Bernoulli-2}) can be separately solved to
determine $h$.




\section{Analytic Similarity Solution}
\label{sec:similarity-solution}

\subsection{A similarity solution modeling disk-jet structure}

\begin{figure}
\begin{center}
\includegraphics[scale=0.8]{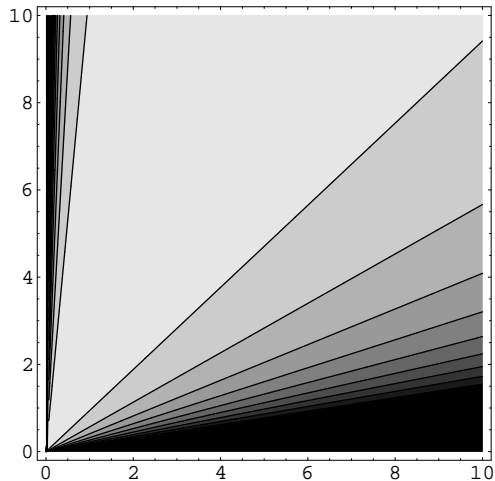}
\caption{ The momentum field (contours of $\psi$ that describe the
streamlines of the poloidal component of $\bm{P}$) of the similarity solution (with $D=1$,
$p=1$, $J=0.1$ and $q=1$). }
\label{fig:flow}
\end{center}
\end{figure}

In this section, we construct a \emph{similarity solution} of the
un-magnetized ($\mu=0$) model (\ref{Beltrami3'}), which describes a
fundamental disk-jet structure.
We define
\begin{equation}
\tau \equiv \frac{z}{r}
\quad (r>0),
\label{similarity-0}
\end{equation}
and an orthogonal variable ($\nabla\tau\cdot\nabla\sigma=0$)
\begin{equation}
\sigma \equiv \sqrt{r^2 + z^2} \ . \label{similarity-0'}
\end{equation}
We consider $\psi$ such that
\begin{equation}
\psi = \psi(\tau) = -J \tau^p  -D \tau^{-q},
\label{similarity-1}
\end{equation}
where $J$ and $p$ ($D$ and $q$) are positive constants, which
control the strength of the jet (disk) flow. As shown in
Fig.\,\ref{fig:flow}, this $\psi$ has a disk-jet-like geometry.

The level sets of $\psi$ (hence, those of $\tau$) are the
streamlines of $\bm{P}$. On the other hand, $\sigma$ serves as the
coordinate directed parallel to the streamlines. We assume that
$\rho_1$ is written as
\begin{equation}
\rho_1(\tau,\sigma) = \rho_\perp(\tau) \rho_\parallel(\sigma),
\label{similarity-1'}
\end{equation}
and, then, $\log\rho_1= \log\rho_\perp(\tau)+\log\rho_\parallel(\sigma)$.

Let us see how the stream function $\psi$ of equation\,(\ref{similarity-1})
satisfies equations (\ref{Beltrami3'}), (\ref{Bernoulli-2})
and (\ref{friction-balance-2}), i.e., we determine all other
fields $I_2(\psi)$, $\rho_1$, $\rho_2$, $\nu$ and $h$ that allow
this $\psi$ to be the solution. For arbitrary $f(\tau)$ and
$g(\tau)$, we observe
\[
{\cal L} f =
\frac{1}{r^2} \left[ (\tau^2+1) f'' + 3\tau f' \right] ,
\]
\[
\nabla f \cdot \nabla g
= \frac{1}{r^2} (\tau^2+1)f' g' .
\]
Hence, the left-had side of equation\,(\ref{Beltrami3'}) is (denoting
$g(\tau)\equiv\log\rho_\perp(\tau)$)
\begin{eqnarray}
& & {\cal L} \psi - \nabla \psi \cdot \nabla \log \rho_1
\nonumber \\
& & ~~= \frac{1}{r^2} \left[ (\tau^2+1) \psi'' + 3\tau \psi' - (\tau^2+1)g'\psi' \right].
\label{similarity-2}
\end{eqnarray}
For this quantity to balance with the right-hand side of
equation\,(\ref{Beltrami3'}), which has no explicit dependence on
$r^{-2}$, both sides must be zero, i.e., we have to set
$I_2'(\psi) =0$ (the implication of this simple condition will be
discussed later). Then, equation\,(\ref{Beltrami3'}) reduces to
\begin{equation}
(\tau^2+1) \psi'' + 3\tau \psi' - (\tau^2+1)g'\psi' = 0.
\label{Beltrami3''}
\end{equation}
We note that the Beltrami condition (\ref{Beltrami3''}) is freed
from $\rho_\parallel(\sigma)$. This fact merits in solving
equation\,(\ref{friction-balance-2}); see equation\,(\ref{friction-balance-3}).

\begin{figure}
\begin{center}
\includegraphics[scale=0.8]{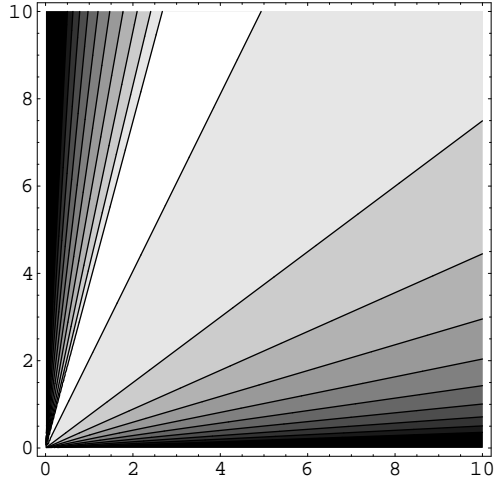}
\caption{ The distribution of $\rho_{\perp}$ of the similarity
solution (with $D=1$, $p=1$, $J=0.1$ and $q=1$). Contour curves
are given in the log scale of $\rho_{\perp}$.
}
\label{fig:density_1}
\end{center}
\end{figure}

For the specific form (\ref{similarity-1}) of $\psi$, we have to
determine an appropriate $g=\log \rho_\perp$ to satisfy
equation\,(\ref{Beltrami3''}), i.e.,
\begin{eqnarray}
g' &=&
\frac{\psi''}{\psi'} + \frac{3\tau}{\tau^2 + 1}
\nonumber
\\
&=&  \frac{J\,p\,(p-1)\tau^{p+q} +
D\,q\,(q+1)}{J\,p\,\tau^{p+q+1}-D\,q\,\tau}
+\frac{3\tau}{\tau^2+1} . \label{similarity-3}
\end{eqnarray}
Solving equation\,(\ref{similarity-3}), we obtain
\[
g \equiv \log\rho_\perp =
\log\frac{|Jp\tau^{p+q}-Dq|}{\tau^{q+1}}
+\frac{3}{2}\log(\tau^2+1) ,
\]
and, thus,
\begin{equation}
\rho_\perp = \frac{(\tau^2+1)^{3/2}|Jp\tau^{p+q}-Dq|}{\tau^{q+1}} .
\label{similarity-4}
\end{equation}
In Fig.\,\ref{fig:density_1}, we show the profile of $\rho_\perp(\tau)$.

\subsection{Bernoulli relation in the disk region}

\begin{figure}
\begin{center}
\includegraphics[scale=0.8]{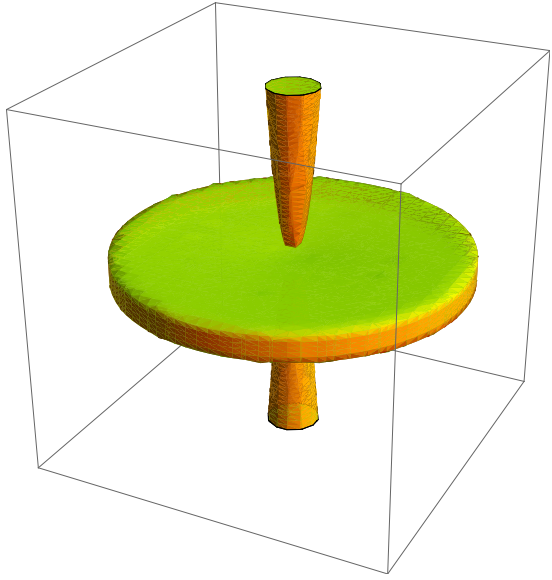}
\caption{ The density $\rho$ in the similarity solution (with
$D=1$, $p=1$, $J=0.1$ and $q=1$). We assume
$\rho_{\parallel}(\sigma)\approx z^{-2/3}$ and $\rho_2\approx
r^{-1/2}$. A levelset surface of $\rho$ is shown in the doiman $r<5$
and $|z|<5$.
}
\label{fig:density}
\end{center}
\end{figure}

As mentioned above, this solution assumes $I_2'(\psi)
~(=\lambda)=0$, and hence, $I_2 = \rho_2 r V_\theta$ must
uniformly distribute. In the disk region (the vicinity of $z=0$),
we may approximate $V_\theta \approx \sqrt{MG/r}$ (Keplerian
velocity). Hence,  $\rho_2 \propto r^{-1/2}$.
In Fig.\,\ref{fig:density}, we show the profile of $\rho=\rho_1\rho_2$
for the case of $\rho_1\propto\rho_\perp$ (i.e., $\rho_\parallel=$constant).

For $\rho=\rho_1\rho_2=\rho_\parallel \rho_\perp r^{-1/2}$,
equation\,(\ref{friction-balance-2}) reads as
\begin{equation}
\nu = \frac{-d\psi/dz}{2 \rho_\parallel \rho_\perp} r^{-3/2}
\propto \frac{r^{-5/2}}{\rho_\parallel(r)}
\label{friction-balance-3}
\end{equation}
along each streamline in the disk region. For a given $\nu$, we
can solve equation\,(\ref{friction-balance-3}) for $\rho_\parallel$ to
determine the density profile.

In the disk region the Bernoulli relation (\ref{Bernoulli-1})
accounts as follows: by $\nabla\cdot\bm{P}=0$, we have $P_r =
\rho V_r \propto r^{-1}$.
If $\rho_\parallel(\sigma) =$ constant, for example,
$\rho\propto r^{-1/2}$ (evaluated along a streamline in the disk region).
Then, we have $V_r = V_{r0} r^{-1/2}$ with a (negative) constant $V_{r0}$.  Combining
the azimuthal velocity $V_\theta = V_{\theta 0} r^{-1/2}$ (which
must be slightly smaller than the Keplerian velocity
$\sqrt{MG/r}$), we obtain
\[
V^2 = (V_{r0}^2 + V_{\theta0}^2) r^{-1} = V_0^2 r^{-1}.
\]
By $\rho_2 \propto r^{-1/2}$, we obtain $ \partial_r (\log \rho_2)
= -(1/2) r^{-1}$. Hence, the Bernoulli relation
(\ref{Bernoulli-1}) demands
\[
\partial_r h = (V_{0}^2-MG) r^{-2},
\]
which yields $h = (MG -V_{0}^2) r^{-1}$. In this estimate, all
components of the energy density (gravitational potential $\phi$,
kinetic energy $V^2/2$ and enthalpy $h$) have a similar profile
($\propto r^{-1}$).

\subsection{Bernoulli relation in the jet region}
In the jet region (vicinity of $r=0$),
the streamlines (contours of $\tau=z/r$) are almost vertical, and
we may approximate $\sigma \approx z$.

Let us first estimate $\rho_2$ using
equation\,(\ref{friction-balance-2}), which is approximated, in the jet
region, by
\begin{eqnarray}
r \rho \nu &=& \{\log\rho_2,\psi\}
\nonumber \\
&\approx&
(\partial_r\psi)\,(\partial_z\log\rho_2)
\nonumber \\
&=& J\,p\,\tau^{p+1}\frac{1}{z}\,
\partial_z(\log\rho_2), \label{jet-Bernoulli-1}
\end{eqnarray}
which shows that $\rho_2$ is an increasing function of $|z|$.
Using $\rho = \rho_\parallel(\sigma)\rho_\perp(\tau)\rho_2$, we
integrate equation\,(\ref{jet-Bernoulli-1}) along the streamline
($\tau=$constant, $\sigma \approx z$):
\begin{equation}
\frac{d\rho_2}{\rho_2^2} = - d\left(\frac{1}{\rho_2}\right)
=\frac{\nu}{Jp}\ \tau^{p+2}\,\rho_\perp(\tau) \rho_\parallel(z) z^2
dz . \label{jet-Bernoulli-2}
\end{equation}
With this $\rho_2(z)$, we may estimate the toroidal (azimuthal)
component of the velocity: $V_\theta = I_2/(r\rho_2) =
(I_2\tau)/(z\rho_2)$, where $I_2$ and $\tau$ are constant (the
latter is constant along each streamline). We find that the
kinetic energy $V_\theta^2/2$ of the azimuthal velocity decreases
as a function of $|z|$ (both by the geometric expansion factor
$z^{-2}$ and the friction damping effect $\rho_2^{-2}$). The steep
gradient of the corresponding hydrodynamic pressure yields a
strong boost near the foot point ($z\approx 0$).

The poloidal component of the kinetic energy is estimated as follows:
We may approximate
\begin{eqnarray}
\frac{1}{2}(V_r^2 + V_z^2) &=& \frac{1}{2\rho^2 r^2}\,|\nabla\psi|^2
\nonumber \\
&\approx& \frac{1}{2\rho^2 r^2} \left(
Jp\frac{z^p}{r^{p+1}}\right)^2
\nonumber \\
&=&
\frac{(Jp)^2\tau^{2p+4}}{2\rho^2z^4} . \label{jet-Bernoulli-2'}
\end{eqnarray}
Here, the vertical distribution of the density $\rho =
\rho_\perp(\tau) \rho_\parallel(\sigma) \rho_2$ is primarily
dominated by $\rho_\parallel(\sigma)\approx\rho_\parallel(z)$.

At long distance from the origin, the jet has a natural similarity
property. For simplicity, let us ignore the effect of the friction
($\nu=0$), and assume $\rho_2=1$. Then, $\rho_\parallel \propto
|z|^{-3/2}$ yields $(V_r^2 + V_z^2)/2 \propto z^{-1}$, which may
balance with the gravitational potential energy $\phi = -MG
|z|^{-1}$. Note, that the azimuthal component of the kinetic
energy disappears at large scale ($V_\theta^2 \propto z^{-2}$).
The Bernoulli condition (\ref{Bernoulli-1}) gives $h$ that also
has a similar distribution of $\propto |z|^{-1}$.

In Fig.\,\ref{fig:density}, we show the profile of $\rho =
\rho_\perp\rho_\parallel\rho_2$ with $\rho_\parallel \propto
|z|^{-3/2}$ (jet region) and $\rho_2 \propto r^{-1/2}$ (disk
region).

\section{Summary and Concluding Remarks}

We have shown that the combination of a \emph{thin} disk and
\emph{narrowly-collimated} jet is the unique structure that is
amenable to the singularity of the Keplerian vorticity; the
\emph{Beltrami} condition, imposing the \emph{alignment} of flow
and \emph{generalized vorticity}, characterizes such a geometry.
Here the conventional vorticity is generalized to combine with magnetic
field (or, electromagnetic vorticity) as well as to subtract the
friction force causing the accretion.

In Section\,\ref{sec:similarity-solution}, we have given an
analytic solution in which the \emph{generalized vorticity} is
purely kinematic ($\bm{\Omega}=\nabla\times\bm{P}_2$ with the
momentum $\bm{P}_2$ that is modified by the friction effect). The
principal force that ejects the jet is, then, the hydrodynamic
pressure dominated by $V_\theta^2/2$. Additional magnetic force
may contribute to jet acceleration if the 
self-consistently generated large-scale magnetic field is
sufficiently large
\cite{bib:Bgen,bib:Bgen2,bib:Bgen3,bib:Bgen4,bib:Bgen5}. Such
structures can be described by the generalized model of
Subsection\,\ref{subsec:general2D}.

The similarity solution has a singularity at the origin (where the
model gravity $\phi=-MGr^{-1}$ is singular), which disconnects the
disk part (parameterized by $D$ and $q$) and the jet part
(parameterized by $J$ and $p$). To ``connect'' both subsystems
(or, to relate $D$, $q$, $J$, and $p$), we need a \emph{singular
perturbation} that dominates the small-scale
hierarchy\,\cite{bib:yoshida2004}. The connection point must
switch the topology of the flow: In the poloidal cross section,
the streamlines of the mass flow, which connect the accreting
inflow and jet's outflow, describe hyperbolic curves converging to
the radial and the vertical axes. On the other hand, the vorticity
lines (or generalized magnetic field lines) thread the disk. This
topological difference of these two vector fields demands
``decoupling'' of them. Instead of disconnecting them by a
singularity, we will have to consider a small-scale structure in
which the topological switch can occur\,\cite{bib:Shiraishi}. The
Hall effect (scaled by $\epsilon$) or the viscosity/resistivity
(scaled by reciprocal Reynolds number) yields a singular
perturbation (then, the generalized \emph{magneto-Bernoulli}
mechanism \cite{bib:mnsy,bib:MSaccel} may effectively accelerate
the jet-flow). In a weakly ionized plasma, the electron equation
(\ref{momentum-e-1}) is modified so that the Hall effect is
magnified by the ratio of the neutral and electron densities; the
ambipoler diffusion effect also yields a higher-order
perturbation\,\cite{bib:krishan2006}. The mechanism of singular
perturbation and the local structure of the disk-jet connection
point may differ depending on the plasma condition near the
central object: in AGN, the plasma is fully ionized but rather
collisional (we also need a relativistic equation of state with
possible presence of pairs), and, in YSO, the plasma is partially
ionized; the relevant dissipation mechanisms determine the scale
hierarchy in the vicinity of the singularity. The study of
above concrete cases was not the scope of present paper and will
be considered elsewhere. Our goal was to show how the 
alignment of flow and \emph{generalized vorticity} condition
arises and how it determines the \emph{singular} structure of a
\emph{thin} disk and \emph{narrowly-collimated} jet invoking the
simplest (minimum) model of magnetohydrodynamics.


\section*{Acknowledgments}

The authors are thankful to Professor R. Matsumoto and Professor G. Bodo for their
discussions and valuable comments. The authors also appreciate discussions with
Professor S. M. Mahajan and Professor V. I. Berezhiani.
This work was initiated at Abdus Salam
International Center for Theoretical Physics, Trieste, Italy.
NLS is grateful for the hospitality
of Plasma Physics Laboratory of Graduate School of Frontier
Sciences at the University of Tokyo during her short term visit in
2007. Work of NLS was partially supported by the Georgian National
Foundation Grant projects 69/07 (GNSF/ST06/4-057) and 1-4/16
(GNSF/ST09-305-4-140).




\section*{References}

\begin{thebibliography}{}


\bibitem{bib:anderson2}
Anderson J M, Li Z Y, Krasnopolsky R and Blandford R D 2005 The
Structure of Magnetocentrifugal Winds. I. Steady Mass Loading {\it
ApJ} {630} 945

\bibitem{bib:BalbusHawley}
Balbus S A and Hawley J F 1998 Instability, turbulence, and
enhanced transport in accretion disks {\it Rev. Mod. Phys.} 70 1




\bibitem{bib:begelman3}
Begelman M C, Blandford R D and Rees M J 1984 Theory of
extragalactic radio sources {\it Rev. Mod. Phys.} 56 255

\bibitem{bib:begelman4}
Begelman M C 1993 Conference summary {\it Astrophysical Jets} \ ed
D Burgarella et al (Cambridge: Cambridge Univ. Press) pp. 305-315

\bibitem{bib:begelman5}
Begelman M C 1998 Instability of Toroidal Magnetic Field in Jets
and Plerions {\it ApJ} 493 291

\bibitem{bib:bland}
Blandford R D and Rees M J 1974 A 'twin-exhaust' model for double
radio sources {\it MNRAS} 169 395

\bibitem{bib:bland1}
Blandford R D and Znajek R L 1977 Electromagnetic extraction of
energy from Kerr black holes {\it MNRAS} 179 433

\bibitem{bib:bland1-2}
Blandford R D and Payne D G 1982 Hydromagnetic flows from
accretion discs and the production of radio jets {\it MNRAS} 199
883

\bibitem{bib:bland2}
Blandford R D 1994 Particle acceleration mechanisms {\it ApJS} 90
515


\bibitem{bib:bogov}
Bogovalov S V and Kelner S R 2010 Accretion and Plasma Outflow
from Dissipationless Discs {\it IJMP D} 19 339

\bibitem{bib:chandrasekhar}
Chandrasekhar S 1956 On Force-Free Magnetic Fields {\it Proc.
Natl. Acad. Sci. USA} 42 1

\bibitem{bib:bland4}
Celotti A and Blandford R D 2001 Black Holes in Binaries and
Galactic Nuclei: Diagnostics, Demography and Formation {\it ESO
Astrophysics Symposia} \  ed L Kaper et al
(Berlin, Heidelberg: Springer-Verlag), 206

\bibitem{bib:ferrari}
Ferrari A 1998 Modeling Extragalactic Jets {\it Annu. Rev. Astron.
Astrophys.} 36 539

\bibitem{bib:ferreira}
Ferreira J 1997 Magnetically-driven jets from Keplerian accretion
discs {\it A\&A} 319 340

\bibitem{bib:ferreira2}
Ferreira J, Dougados C and Cabrit S 2006 Which jet launching
mechanism(s) in T Tauri stars? {\it A\&A} 453 785

\bibitem{bib:ferreira3}
Ferreira J, Dougados C and Whelan E 2007 Jets from Young Stars I:
Models and Constraints {\it Lecture Notes in Physics} \  ed J
Ferreira et al (Berlin, Heidelberg: Springer-Verlag) 723 181

\bibitem{bib:Hartigan}
Hartigan P, Edwards S and Ghandour L 1995 Disk Accretion and Mass
Loss from Young Stars {\it ApJ} 452 736

\bibitem{bib:Bgen}
Hawley J F, Gammie C F and Balbus S A 1995 Local Three-dimensional
Magnetohydrodynamic Simulations of Accretion Disks {\it ApJ} 440
742



\bibitem{bib:begelman-pressure}
Heinz S and Begelman M C 2000 Jet Acceleration by Tangled Magnetic
Fields {\it ApJ} 535 104

\bibitem{bib:Jones}
Jones D L, Werhle A E, Meier D L and Piner B G 2000 The Radio Jets
and Accretion Disk in NGC 4261 {\it ApJ} 534 165

\bibitem{bib:magn}
K\"onigl A and Pudritz R E 2000 Disk Winds and the
Accretion-Outflow Connection {\it Protostars and Planets IV} \ ed
V Mannings et al (Tuscon: Univ. Arizona Press) 759

\bibitem{bib:bland3}
Krasnopolsky R, Li Z Y and Blandford R D 1999 Magnetocentrifugal
Launching of Jets from Accretion Disks. I. Cold Axisymmetric Flows
{\it ApJ} 526 631

\bibitem{bib:bland5}
Krasnopolsky R, Li Z Y and Blandford R D 2003 Magnetocentrifugal
Launching of Jets from Accretion Disks II. Inner Disk-driven Winds
{\it ApJ} 595 631

\bibitem{bib:krishan2006}
Krishan V and Yoshida Z 2009 Kolmogorov dissipation scales in
weakly ionized plasmas {\it MNRAS} 395 2039

\bibitem{bib:shibata}
Kudoh T and Shibata K 1997 Magnetically Driven Jets from Accretion
Disks. I. Steady Solutions and Application to Jets/Winds in Young
Stellar Objects {\it ApJ} 474 362

\bibitem{bib:matsumoto2}
Kudoh T, Matsumoto R and Shibata K 2003 MHD simulations of jets
from accretion disks {\it Astrophys. Sp. Sci.} 287 99

\bibitem{bib:shibata2}
Kuwabara T, Shibata K, Kudoh T and Matsumoto R 2005 The
Acceleration Mechanism of Resistive Magnetohydrodynamic Jets
Launched from Accretion Disks {\it ApJ} 621 921


\bibitem{bib:livio}
Livio M 1997 The Formation Of Astrophysical Jets {\it Accretion
Phenomena and Related Outflows; IAU Colloquium 163} \ ed. D T
Wickramasinghe et al (San Francisco: ASP) {\it ASP Conference
Series} 121 845

\bibitem{bib:livio2}
Livio M., 1999, Phys. Rep., 311, 225

\bibitem{bib:low}
Low B C 1982 Nonlinear force-free magnetic fields {\it Rev.
Geophys. Space Phys.} 20 145

\bibitem{bib:matsumoto}
Machida M and Matsumoto R 2003 Global Three-dimensional
Magnetohydrodynamic Simulations of Black Hole Accretion Disks:
X-Ray Flares in the Plunging Region {\it ApJ} 585 429

\bibitem{bib:beltrami}
Mahajan S M and Yoshida Z 1998 Double Curl Beltrami Flow:
Diamagnetic Structures {\it Phys. Rev. Lett.} 81 4863

\bibitem{bib:mnsy}
Mahajan S M, Nikol'skaya K I, Shatashvili N L and Yoshida Z 2002
Generation of Flows in the Solar Atmosphere Due to Magnetofluid
Coupling {\it ApJ} 576 L161

\bibitem{bib:MSaccel}
Mahajan S M, Shatashvili N L, Mikeladze S V and Sigua K I 2006
Acceleration of plasma flows in the closed magnetic fields:
Simulation and analysis {\it Phys. Plasmas} 13 062902


\bibitem{bib:Bgen2}
Matsumoto R and Tajima T 1995 Magnetic viscosity by localized
shear flow instability in magnetized  {\it ApJ} 445 767

\bibitem{bib:matsumoto3}
Matsumoto R, Machida M and Nakamura K 2004 Global 3D MHD
Simulations of Optically Thin Black Hole Accretion Disks {\it
Prog. Theor. Phys. Suppl.} 155 124

\bibitem{bib:Bgen3}
Ogilvie G I and Livio M 1998 On the Difficulty of Launching an
Outflow from an Accretion Disk {\it ApJ} 499 329


\bibitem{bib:pellet}
Pelletier G and Pudritz R E 1992 Hydromagnetic disk winds in young
stellar objects and active galactic nuclei {\it ApJ} 394 117


\bibitem{bib:shakura}
Shakura N I and Sunyaev R 1973 Black holes in binary systems.
Observational appearance {\it A\&A} 24 337

\bibitem{bib:Shiraishi}
Shiraishi J, Yoshida Z and Furukawa M 2009 Topological Transition
from Accretion to Ejection in a Disk-Jet System -— Singular
Perturbation of the Hall Effect in a Weakly Ionized Plasma {\it
ApJ} 697 100


\bibitem{bib:Bgen4}
Tout C A and Pringle J E 1996 Can a disc dynamo generate
large-scale magnetic fields? {\it MNRAS} 281 219

\bibitem{bib:Bgen5}
Vlemmings W H T, Bignall H E and Diamond P J 2007 Green Bank
Telescope Observations of the Water Masers of NGC 3079: Accretion
Disk Magnetic Field and Maser Scintillation {\it ApJ} 656 198

\bibitem{bib:yoshida-giga}
Yoshida Z and Giga Y 1990 Remarks on spectra of operator ROT {\it
Math. Z.} 204 235



\bibitem{bib:yoshida2004}
Yoshida Z, Mahajan S M and Ohsaki S 2004 Scale hierarchy created
in plasma flow {\it Phys. Plasmas} 11 3660

\bibitem{bib:yoshida2009}
Yoshida Z 2009 Clebsch parameterization: Basic properties and
remarks on its applications {\it J. Math. Phys.} 50 113101

\bibitem{bib:zanni}
Zanni C, Ferrari A, Rosner R, Bodo G and Massaglia S 2007 MHD
simulations of jet acceleration from Keplerian accretion disks.
The effects of disk resistivity {\it A\&A} 469 811


\end{thebibliography}
\end{document}